\documentclass[preprint,prb,showpacs,showkeys]{revtex4}%
\usepackage{amsfonts}
\usepackage{amsmath}
\usepackage{amssymb}
\usepackage{graphicx}%
\begin{document}
\preprint{ }
\title{Molecular Theory of Irreversibility}
\author{A. P\'{e}rez-Madrid}
\affiliation{Departament de F\'{\i}sica Fonamental, Facultat de F\'{\i}sica, Universitat de
Barcelona, Diagonal 647, 08028 Barcelona, SPAIN}

\begin{abstract}
A generalization of the Gibbs entropy postulate is proposed based on the BBGKY
[Bogolyubov-Born-Green-Kirkwood-Yvon] hierarchy of equations as the
nonequilibrium entropy for a system of N interacting particles. This entropy
satisfies the basic principles of \ thermodynamics in the sense that it
reaches its maximum at equilibrium and is coherent with the second law. By
using a generalization of the Liouville equation describing the evolution of
the distribution vector, it is demonstrated that the entropy production is a
non-negative quantity. Moreover, following the procedure of non-equilibrium
thermodynamics a transport matrix is introduced and a microscopic expression
for this is derived. This framework allows one to perform the thermodynamic
analysis of non-equilibrium steady states with smooth phase space distribution
functions which, as proven here, constitute the states of minimum entropy
production when one considers small departures from stationarity.

\end{abstract}
\date[]{}
\maketitle

\section{Introduction}

According to the mechanicistic interpretation of the physical world, the basic
laws of nature are deterministic and time reversible. However, at the
macroscopic level we observe irreversible processes related to energy
degradation which generates entropy. How do we reconcile the `spontaneous
production of entropy' with the time reversibility of the microscopic
equations of motion?. At the end of the nineteenth century, Boltzmann tried to
answer this question from a probabilistic point of view. According to him,
entropy is a measure of the lack of knowledge of the precise state of matter
and can be defined as a function of the probability of a given state of
matter. This function associates a number $S_{B}(X)=\log\left\vert
\Gamma_{M(X)}\right\vert $ to each microstate $X$\ of a macroscopic system,
with $\left\vert \Gamma_{M}\right\vert $\ being the volume of the region of
the phase space $\Gamma_{M}$\ corresponding to the macrostate $M=M(X)$. The
macrostate $M$ is all of a group of states $Y$ such that $M(Y)=M(X)=M$. In
this sense, the Boltzmann entropy is a function of the microstate which at
equilibrium coincides with the thermodynamic entropy. All systems in their
irreversible evolution tend to a state of maximum probability or maximum
entropy -the state of equilibrium-.

In contrast to the Boltzmann entropy, the Gibbs entropy is not a function of
the individual microstate but rather a function of the probability
distribution in a statistical ensemble of systems with both entropies
coinciding at equilibrium. As a consequence of the incompressible character of
the flow of points representing the natural evolution of the statistical
ensemble in phase space , the Gibbs entropy is a constant of motion. Thus, it
has been argued that the relevant entropy for understanding thermodynamic
irreversibility is the Boltzmann entropy and not the Gibbs entropy
\cite{lebowitz2}$^{-}$ \cite{goldstein}.

In addition, the problem of the diverging character of the Gibbs entropy
related to the negative sign of the entropy production in non-equilibrium
stationary states apparently excludes the use of the Gibbs entropy in the
statistical description of non-equilibrium systems \cite{andrey}$^{,}$
\cite{hoover}. This raises the question as to how to define the
non-equilibrium entropy and if possible, to give a thermodynamic description
of non-equilibrium fluctuations. In other words: can Thermodynamics describe
systems far from equilibrium? \cite{gallavotti}$^{,}$ \cite{ruelle0}.

Thus, from the moment when Gibbs first postulated his entropy formula, the
definition of the non-equilibrium entropy and its relation to irreversibility
has been an outstanding problem, now compounded by the fact that the entropy
production is negative in non-equilibrium stationary states in apparent
violation of the second law of \ Thermodynamics \cite{evans}$^{,}$
\cite{wang}. This constitutes an open problem which must be solved.

A huge amount of work has been done on this subject trying to address the
problem. On one hand, there have been attempts to extend the equilibrium
entropy \cite{cohen}$^{-}$\cite{gaspard} to non-equilibrium situations in
order to avoid the divergence of the Gibbs entropy. On the other hand, work
has been done to establish fluctuation theorems for the probability of the
entropy production fluctuations \cite{evans2}. In a previous work
\cite{agusti} we showed a way to circumvent the difficulty of reconciling the
second law of thermodynamics with the reversible microscopic equations of
motion in the framework of the Bogolyubov-Born-Green-Kirkwood-Yvon (BBGKY)
hierarchy. We proposed a functional of the set of \ s-particle reduced
distribution functions as the entropy for a system of N interacting particles.
This entropy does not enter into contradiction with thermodynamics, and as we
show here, in addition to being time-dependent it enables one to perform a
thermodynamic analysis of the stationary non-equilibrium states. In this
sense, our theory constitutes an extension of the scope of thermodynamics to
systems away from equilibrium.

We begin this contribution introducing in section 2 the representation of the
state of an isolated system in terms of the hierarchy of reduced distribution
functions. Afterwards, in section 3, we introduce a generalization of the
Gibbs entropy postulate as the non-equilibrium entropy and derive the entropy
production. After that, in section 4, we give the thermodynamic analysis of
the non-equilibrium steady states of the system and establish a minimum
principle for those states having obtained a phenomenological transport matrix
previously. Finally, in section 5, we stress our main conclusions.

\section{Hamiltonian Dynamics}

Let's consider a dynamical system of N identical particles whose Hamiltonian
$H_{N}(\left\{  \underline{q}^{N},\underline{p}^{N}\right\}  )$ is given by
\begin{equation}
H_{N}=\sum_{j=1}^{N}\frac{\underline{p}_{j}^{2}}{2m}+\frac{1}{2}\sum_{j\neq
k=1}^{N}\phi\left(  \left\vert \underline{q}_{j}-\underline{q}_{k}\right\vert
\right)  \text{ ,}\label{hamiltonian}%
\end{equation}
where $m$ is the mass and $\phi\left(  \left\vert \underline{q}_{j}%
-\underline{q}_{k}\right\vert \right)  \equiv\phi_{jk}$ is the interaction
potential. The state of the system is completely specified at a given time by
the N-particle distribution function $F\left(  \left\{  \underline{q}%
^{N},\underline{p}^{N}\right\}  ;t\right)  $ which evolves in time according
to the Liouville equation. Nonetheless, an strictly equivalent alternative
description of the state of the system can be given in terms of the
distribution vector \cite{balescu}
\begin{equation}
\underline{f}\equiv\left\{  f_{o},f_{1}(\Gamma_{1}),f_{2}(\Gamma
_{2}),.........,f_{N}(\Gamma_{N})\right\}  \text{ ,}\label{distributionvector}%
\end{equation}
with $\Gamma_{s}=(x_{1},x_{2,}......,x_{s})$ and $x_{j}\equiv(\underline
{q}_{j},\underline{p}_{j})$. Additionally, the set of quantities $\Gamma_{s}$
can be grouped as the vector $\underline{\Gamma}\mathbf{\equiv}\left\{
\Gamma_{1},\Gamma_{2},.........,\Gamma_{N}\right\}  $, and correspondingly
$\underline{H}\mathbf{\equiv}\left\{  H_{1},H_{2},.........,H_{N}\right\}  $
can be defined with $H_{s}$ being the s-particle Hamiltonian. The distribution
vector represents the set of all the s-particle reduced distribution functions
$f_{s}(\Gamma_{s})$ $(s=0,....,N)$, defined through
\begin{equation}
f_{s}=\frac{N!}{(N-s)!}\int F(x_{1},..,x_{s},x_{s+1},..,x_{N})\text{ }%
dx_{s+1}...dx_{N}\text{ .}\label{reduceddistribution}%
\end{equation}
The evolution equations of these functions can be obtained by integrating the
Liouville equation thereby constituting a set of coupled equations: the BBGKY
hierarchy which can be written in a compact way as a generalized Liouville
equation \cite{agusti}
\begin{equation}
\left(  \frac{\partial}{\partial t}-\mathcal{PL}\right)  \underline
{f}(t)=\mathcal{QL}\underline{f}(t)\text{ ,}%
\label{separatedgeneralizedliouville}%
\end{equation}
with
\begin{align}
\langle s\left\vert \mathcal{PL}\right\vert s^{\prime}\rangle &
=\delta_{s,s^{\prime}}\left\{  \sum_{j=1}^{s}L_{j}^{o}+\sum_{j<n=1}^{s}%
L_{j,n}^{\prime}\right\}  \text{ }\label{diagonal}\\
&  =\delta_{s,s^{\prime}}\left[  H_{s},...\right]  _{P}\text{,}\nonumber
\end{align}
and
\begin{equation}
\langle s\left\vert \mathcal{QL}\right\vert s^{\prime}\rangle=\delta
_{s^{\prime},s+1}\int\left\{  \sum_{j=1}^{s}L_{j,s+1}^{\prime}\right\}
dx_{s+1}\text{ .}\label{nondiagonal}%
\end{equation}
where $\mid s\rangle$ represents the s-particle state defined through
$\left\langle \Gamma_{s}\right\vert s\rangle=f_{s}(\Gamma_{s})$ and where the
projection operators $\mathcal{P}$ and $\mathcal{Q}$, its complement with
respect to the identity, give the diagonal and non-diagonal part of the
generalized Liouvillian $\mathcal{L}$, respectively. Here, $L_{j}^{o}=\left[
H_{j}^{o},...\right]  _{P}$ ,where $\left[  ...,...\right]  _{P}$ is the
Poisson bracket, $H_{j}^{o}=\frac{\underline{p}_{j}^{2}}{2m}$, and
$L_{j,n}^{\prime}=\left[  H_{j,n}^{\prime},...\right]  _{P}$ , with
$H_{j,n}^{\prime}=\frac{1}{2}\phi_{j,n}$. In the next section we will show
that irreversibility is manifested in the dynamics of the system when the
adequate description, \textit{i.e. }in terms of the distribution vector is used.

\section{Non-equilibrium entropy and irreversibility}

As the expression for the non-equilibrium entropy we propose%
\begin{align}
S  &  =-k_{B}Tr\left\{  \underline{f}\log\left(  \underline{f}_{o}%
^{-1}\underline{f}\right)  \right\}  +S_{o}\nonumber\\
&  =-k_{B}\sum_{n=1}^{N}\frac{1}{n!}\int f_{n}\log\frac{f_{n}}{f_{o,n}%
}\;dx_{1}.....dx_{n}\text{ }+S_{o}\text{ ,} \label{gibbs}%
\end{align}
a functional of $\underline{f}$, generalizing the Gibbs entropy postulate
\cite{degroot}$^{,}$ \cite{vankampen}, based on the fact that the distribution
vector determines the state of the system. The Gibbs entropy postulate is also
known as relative or Kullback entropy\cite{schlogl} $^{,}$\cite{wehrl}. Here,
$k_{B}$ is the Boltzmann constant, $S_{o}$ the equilibrium entropy and
$\underline{f}_{o}$ is assumed to be the equilibrium distribution vector which
corresponds with $S_{o}$, satisfying $\mathcal{L}\underline{f}_{o}=0$, the
Yvon-Born-Green (YBG)\ hierarchy. Moreover, $S$ is maximum at equilibrium when
$\underline{f}\mathbf{=}\underline{f}_{o}$ giving $S=S_{o}$, which can be
proven by taking the first and second variation of $S$ with respect to
$\underline{f}$ while $S_{o}$ and $\underline{f}_{o}$ remain fixed%
\begin{equation}
\delta S=-k_{B}Tr\left\{  \delta\underline{f}\log\left(  \underline{f}%
_{o}^{-1}\underline{f}\right)  \right\}  \label{firstvariation}%
\end{equation}
showing that $\delta S=0$ at equilibrium when $\underline{f}=\underline{f}%
_{o}$ and
\begin{equation}
\delta^{2}S=-\frac{1}{2}k_{B}Tr\left\{  \underline{f}_{o}\left(  \underline
{f}_{o}^{-1}\delta\underline{f}\right)  ^{2}\right\}  \label{secondvariation}%
\end{equation}
which is a negative quantity.

As known the Gibbs entropy is a functional of the full phase-space
distribution function. Since the flow of points representing the natural
evolution of an isolated system in phase space is incompressible, this entropy
is a constant of motion under the standard Liouville dynamics which is not the
case for the generalized Gibbs entropy (\ref{gibbs}). This entropy is coherent
with the second law according to which $S$ increases in irreversible processes
such as the relaxation to equilibrium from an initially non-equilibrium state.
To prove this we compute the rate of change of $S$ which can be obtained by
differentiating Eq. (\ref{gibbs}) with the help of Eq.
(\ref{separatedgeneralizedliouville})
\begin{align}
\frac{dS}{dt}  &  \equiv\sigma=-k_{B}Tr\left\{  \frac{\partial\underline{f}%
}{\partial t}\log\left(  \underline{f}_{o}^{-1}\underline{f}\right)  \right\}
\label{entropyproduction}\\
&  =-k_{B}Tr\left\{  \mathcal{L}\underline{f}\log\left(  \underline{f}%
_{o}^{-1}\underline{f}\right)  \right\}  \geqq0\text{ .}\nonumber
\end{align}
This constitutes the entropy production corresponding to the relaxation
passing from a non-equilibrium state to equilibrium. Two contributions to the
entropy production $\sigma$ (\ref{entropyproduction}) can be distinguished%
\begin{align}
\sigma_{1}  &  =-k_{B}Tr\left\{  \mathcal{PL}\underline{f}\log\left(
\underline{f}_{o}^{-1}\underline{f}\right)  \right\} \nonumber\\
&  =-\frac{1}{T}\sum_{n=1}^{N}\frac{1}{n!}\sum_{j=1}^{n}\int f_{n}%
\underline{p}_{j}\cdot\left(  -k_{B}T\nabla_{j}\log f_{o,n}+\sum_{j\neq
i=1}^{n}\underline{F}_{j,i}\right)  dx_{1}....dx_{n}\text{ ,} \label{sigma1}%
\end{align}
where $T$ is the equilibrium kinetic temperature, taking into account that the
dependence of $f_{o,n}$ in the velocities is given through a local Maxwellian
and $\underline{F}_{j,i}=-\nabla_{j}\phi_{j,i}$. The second contribution is
\begin{align}
\sigma_{2}  &  =-k_{B}Tr\left\{  \mathcal{QL}\underline{f}\log\left(
\underline{f}_{o}^{-1}\underline{f}\right)  \right\} \nonumber\\
&  =k_{B}\sum_{n=1}^{N}\frac{1}{n!}%
{\displaystyle\int}
\left(  \sum_{j=1}^{n}\underline{F}_{j,n+1}\cdot\frac{\partial}{\partial
\underline{p}_{j}}f_{n+1}\right) \nonumber\\
&  \times\log\frac{f_{n}}{f_{o,n}}\;dx_{1}.....dx_{n}dx_{n+1}\text{ \ .}
\label{entropyproduction1}%
\end{align}
After partial integration Eq. (\ref{entropyproduction1}) gives%
\begin{align}
\sigma_{2}=-k_{B}\sum_{n=1}^{N}\frac{1}{n!}%
{\displaystyle\int}
\sum_{j=1}^{n}\left(
{\displaystyle\int}
\underline{F}_{j,n+1}f_{n+1}dx_{n+1}\right)   &  \cdot\frac{\partial}%
{\partial\underline{p}_{j}}\log\frac{f_{n}}{f_{o,n}}\;dx_{1}...dx_{n}%
\nonumber\\
=-k_{B}\sum_{n=1}^{N}\frac{1}{n!}\sum_{j=1}^{n}%
{\displaystyle\int}
f_{n}\underline{\mathcal{F}}_{j}(\underline{q}_{j})  &  \cdot\frac{\partial
}{\partial\underline{p}_{j}}\log\frac{f_{n}}{f_{o,n}}\;dx_{1}...dx_{n}\text{
\ ,} \label{entropyproduction2}%
\end{align}
where $f_{n}\underline{\mathcal{F}}_{j}(\underline{q}_{j})=%
{\displaystyle\int}
\underline{F}_{j,n+1}f_{n+1}dx_{n+1}$. Hence,
\begin{gather}
\sigma_{2}=-k_{B}\sum_{n=1}^{N}\frac{1}{n!}\sum_{j=1}^{n}%
{\displaystyle\int}
\underline{\mathcal{F}}_{j}\cdot\left(  \frac{\partial f_{n}}{\partial
\underline{p}_{j}}-f_{n}\frac{\partial\log f_{o,n}}{\partial\underline{p}_{j}%
}\right)  dx_{1}...dx_{n}\nonumber\\
=-\frac{1}{T}\sum_{n=1}^{N}\frac{1}{n!}\sum_{j=1}^{n}%
{\displaystyle\int}
f_{n}\underline{\mathcal{F}}_{j}\cdot\underline{p}_{j}dx_{1}.....dx_{n}\text{
\ .} \label{entropyproduction3}%
\end{gather}
Therefore,%
\begin{equation}
\sigma=\sigma_{1}+\sigma_{2}\text{ \ .} \label{total sigma}%
\end{equation}
At equilibrium when we substitute $f_{o,n}$ for $f_{n}$, $\sigma_{1}$ and
$\sigma_{2}$ becomes zero, while in any other case these are not necessarily
zero. Hence, the interaction among the different particles causing the
correlations introduces the irreversibility into the system which is
accompanied by a creation of entropy. Entropy should explicitly contain all
the correlations among the particles through the reduced distribution
functions according to Eq. (\ref{gibbs}). This is our most important result in
this paper. Although one can find somewhat related analyses \cite{kirkwood}%
$^{-}$\cite{green}, as far as we know, this expression (\ref{gibbs}) based on
the BBGKY hierarchy has no direct counterpart in the previous literature.

As compared to other definitions of entropy, Eq. (\ref{gibbs}) does not
contain any coarse-graining since this would imply some kind of time-averaging
in order to subsum a certain number of time-scales, a procedure absent in Eq.
(\ref{gibbs}). In addition, the generalized Gibbs entropy (\ref{gibbs}) is not
a constant of motion under the Liouville dynamics given by Eq.
(\ref{separatedgeneralizedliouville}). This dynamics introduces a
trace-preserving transformation which in concomitance with the fact that Eq.
(\ref{gibbs}) is a concave functional of the distribution vector leads to the
positive character of the entropy production. In this framework we have shown
in Ref. \cite{agusti} that by using the methods of the non-equilibrium
\ thermodynamics it is possible to recover the kinetic Boltzmann equation. In
the light of these facts not contained in other definitions of the entropy in
the literature the strength of our definition (\ref{gibbs}) resides. Hence, it
seems that the generalized Gibbs entropy (\ref{gibbs}) built in the framework
of the BBGKY hierarchy which explicitly incorporates the correlations between
all particles clusters in the system constitutes the appropriate description
leading to irreversibility at the macroscopic level.

In appendix A we give the local version constituting the valance equation of
the specific entropy.

\section{Non-equilibrium steady~ \noindent states\noindent}

As a way of demonstrating the applicability of the framework established in
the previous section which enables us to describe irreversibility in a N-body
system, let us assume that a non-conservative field $h(t)$ acts on this
system, modifying the Hamilton equations of motion%
\begin{gather}
\underline{\dot{q}}_{i}=\frac{\partial H_{N}}{\partial\underline{p}_{i}%
}+\underline{C}_{i}(x_{1},...x_{N})h(t)\text{ ,}\nonumber\\
\underline{\dot{p}}_{i}=-\frac{\partial H_{N}}{\partial\underline{q}_{i}%
}+\underline{D}_{i}(x_{1},...x_{N})h(t)\text{ ,} \label{hamilton}%
\end{gather}
where $\underline{C}_{i}(x_{j})$ and $\underline{D}_{i}(x_{j})$ are coupling
functions which are assumed to be analytic. This drive introduces a
compressible contribution to the flow in the phase space which is reflected in
the Liouville equation characterizing the phase space flow%
\begin{equation}
\frac{\partial}{\partial t}F-\left[  H_{N},F\right]  _{P}=-\frac{\partial
}{\partial\Gamma_{N}}\cdot\dot{\Gamma}_{N}^{nc}F\text{ ,}
\label{generalizedliouvil}%
\end{equation}
where $\dot{\Gamma}_{N}^{nc}=(\dot{x}_{1}^{nc},....,\dot{x}_{N}^{nc})$,
$\ \dot{x}_{i}^{nc}=\left(  \underline{C}_{i}(x_{j}),\underline{D}_{i}%
(x_{j})\right)  h(t)$ . Here, the full phase-space distribution function $F$
should be an smooth function in order that Eq. (\ref{gibbs}) be applicable.
This function is defined through%
\begin{equation}
F(\Gamma_{N},t)=\frac{1}{M}%
{\displaystyle\sum_{m=1}^{M}}
\delta\left(  \Gamma_{N}-\gamma_{m}(t)\right)  \text{ \ ,}
\label{distribution}%
\end{equation}
with $\gamma_{m}(t)$ being one among the M realizations $(m=1,......,M)$ in
the statistical mechanical ensemble associated to our N-body system, so that
$MF(\Gamma_{N},t)d\Gamma_{N}$ equals the number of realizations contained in
the infinitesimal phase space volume $d\Gamma_{N}$ at time t\cite{reimann}.

In this context, some simulations have shown multifractal phase-space
distributions \cite{holian}$^{-}$\cite{evans3}. Although the multifractal
nature of the phase space distribution functions has been stated,
\cite{evans4}$^{,}$\cite{evans5}, apparently this is not always the case. In
fact, it has been shown that in the case of the Galton staircase
\cite{tuckerman2}$^{,}$\cite{tuckerman3} an initial strong fractal phase space
portrait becomes smooth when the system is coupled to a more ergodic
Hoover-Holian thermostat. 

To Eq. (\ref{generalizedliouvil}) corresponds
\begin{equation}
\left(  \frac{\partial}{\partial t}-\mathcal{P}\left(  \mathcal{L}%
-\underline{G}\right)  \right)  \underline{f}=\mathcal{QL}\underline{f}\text{
}\label{genralizedliouville2}%
\end{equation}
for the distribution vector, where $\underline{G}\underline{f}\mathbf{=}$
$\frac{\partial}{\partial\underline{\Gamma}}\left(  \underline{\dot{\Gamma}%
}^{nc}\underline{f}\right)  $ and $\langle s^{\prime}|\mathcal{P}\underline
{G}\underline{f}|s\rangle=\delta_{s,s^{\prime}}\frac{\partial}{\partial
\Gamma_{s}}\cdot\left(  \dot{\Gamma}_{s}^{nc}f_{s}\right)  $. \bigskip The
total rate of change of the entropy is given now by%
\begin{equation}
\frac{dS}{dt}=k_{B}Tr\left\{  \underline{G}\underline{f}\log\left(
\underline{f}_{o}^{-1}\underline{f}\right)  \right\}  +\sigma\text{.}%
\label{stationary entropy}%
\end{equation}
For a constant field $h$ a steady state, $dS/dt=0$, can be reached when both
contributions appearing in the right hand side of the previous equation
balance each other. In such a case, the distribution vector is $\underline
{f}_{st}$ and the entropy production (\ref{entropyproduction}) reduces to%
\begin{align}
\sigma_{st} &  =-k_{B}Tr\left\{  \underline{G}\underline{f}_{st}\log\left(
\underline{f}_{o}^{-1}\underline{f}_{st}\right)  \right\}  \nonumber\\
&  =k_{B}Tr\left\{  \underline{f}_{st}\underline{\dot{\Gamma}}^{nc}%
\frac{\partial}{\partial\underline{\Gamma}}\log\left(  \underline{f}_{o}%
^{-1}\underline{f}_{st}\right)  \right\}  \text{ .}%
\label{stationaryproduction}%
\end{align}

In an attempt to study the stability of this non-equilibrium steady state we
come back to Eq. (\ref{entropyproduction}). From this equation we infer the
phenomenological law%
\begin{equation}
\frac{\partial\underline{f}}{\partial t}\equiv-\underline{M}\log\left(
\underline{f}_{o}^{-1}\underline{f}\right)  \label{phenomenological}%
\end{equation}
in analogy with non-equilibrium thermodynamics \cite{degroot}, where
$\underline{M}$ is a transport matrix which in general might depend on the
state of the system. Thus
\begin{equation}
\sigma=k_{B}Tr\left\{  \log\left(  \underline{f}_{o}^{-1}\underline{f}\right)
\underline{M}\log\left(  \underline{f}_{o}^{-1}\underline{f}\right)  \right\}
\text{ ,}\label{quadraticform}%
\end{equation}
expressing $\sigma$ as a positive semidefinite quadratic form. Therefore, in
light of our result corresponding to Eq. (\ref{entropyproduction}) one can
state that $\underline{M}$ is a positive semidefinite matrix which in addition
should be Hermitian%
\begin{equation}
M\mathbf{(}s\mathbf{\mid}s^{\prime})=M^{\dagger}\mathbf{(}s\mathbf{\mid
}s^{\prime})=\text{ }M^{c}\mathbf{(}s^{\prime}\mathbf{\mid}s)\text{
,}\label{hermiticity}%
\end{equation}
as predicted by the Onsager symmetry relations. Here $^{\dagger}$ refers to
the Hermitian conjugate and $^{c}$ stands for the complex conjugated. By
comparing Eqs. (\ref{total sigma}) and (\ref{quadraticform}) we can obtain a
microscopic expression for $\underline{M}$%
\begin{gather}
Tr\left\{  \log\left(  \underline{f}_{o}^{-1}\underline{f}\right)
\underline{M}\log\left(  \underline{f}_{o}^{-1}\underline{f}\right)  \right\}
\nonumber\\
=\sigma_{1}+\sigma_{2}\text{ \ ,}\label{green-kubo}%
\end{gather}
where $\sigma_{1}$and $\sigma_{2}$ are given by Eqs (\ref{sigma1}) and
(\ref{entropyproduction3}), respectively. In order to get a Lyapounov function
for this system we can develop the expression (\ref{quadraticform}) as follows%
\begin{gather}
\sigma=k_{B}Tr\left\{  \log\left(  \underline{f}_{st}^{-1}\underline
{f}\right)  \underline{M}\log\left(  \underline{f}_{st}^{-1}\underline
{f}\right)  \right\}  \nonumber\\
+2k_{B}Tr\left\{  \log\left(  \underline{f}_{st}^{-1}\underline{f}\right)
\underline{M}\log\left(  \underline{f}_{o}^{-1}\underline{f}_{st}\right)
\right\}  \nonumber\\
+k_{B}Tr\left\{  \log\left(  \underline{f}_{o}^{-1}\underline{f}_{st}\right)
\underline{M}\log\left(  \underline{f}_{o}^{-1}\underline{f}_{st}\right)
\right\}  \text{ .}\label{intermidiate}%
\end{gather}
Notice that the second and third terms on the right hand side of this equation
vanishes in view to the fact that $\underline{M}\log\left(  \underline{f}%
_{o}^{-1}\underline{f}_{st}\right)  =0$ according to Eq.
(\ref{phenomenological}). On the other hand, for small departures from the
steady state we can assume that $\underline{M}$ coincides with its value at
the steady state. Thus, the remaining non-zero term in Eq. (\ref{intermidiate}%
) provides us the desired Lyapounov function
\begin{equation}
L(t)\equiv k_{B}Tr\left\{  \log\left(  \underline{f}_{st}^{-1}\underline
{f}\right)  \underline{M}\log\left(  \underline{f}_{st}^{-1}\underline
{f}\right)  \right\}  \label{lyapunov}%
\end{equation}
which is a positive semidefinite quadratic form having its minimum value at
the stationary state $\underline{f}_{st}$. Hence, the steady state is the
state of minimum entropy production. This is a result analogous to the one of
the principle of minimum entropy production by I. Prigogine \cite{prigogine},
nonetheless, the later was formulated in the framework of the classical
non-equilibrium thermodynamics applicable in the linear regime.

\section{Conclusions}

Here we have shown a representation of the statistical description of a
many-body system in terms of the distribution vector which reveals the
irreversible manifestation of the motion of this system at the macroscopic
level. This manifestation is hidden when one works with the standard Liouville
equation but at the level of the generalized Liouville equation this
irreversibility becomes evident. The Liouville equation is a closed equation
for the phase space distribution function while the generalized Liouville
equation encloses a set of coupled equations, each one in itself representing
a contraction of the statistical description. But, as I have mentioned
previously, the full statistical description in terms of the phase space
distribution function is entirely equivalent to the description in terms of
the set of all the reduced s-particle distribution functions, the distribution
vector, with no loss of information. In this scenario, we show that the
entropy given through the proposed generalization of the Gibbs entropy
postulate is not conserved throughout the motion in phase space. The key point
is that our entropy contains all the correlations among the particles through
the reduced distribution functions. We have proven that this entropy reaches
its maximum value at the equilibrium state where it coincides with the
thermodynamic entropy and that the entropy production is positive according to
the second law.

By applying the methods of \ non-equilibrium thermodynamics in phase space,
\textit{i.e.} in the framework of the mesoscopic non-equilibrium
thermodynamics MNET , a theory which has been proved to be successful in
deriving kinetic transport equations in some particular cases\cite{agusti2}%
$^{-}$\cite{rubi}, we have obtained a phenomenological transport matrix
relating the rate of change of the distribution vector with its conjugated
thermodynamic force. We have proven that this is a positive semidefinite
Hermitian matrix satisfying a kind of Green-Kubo relation. In this scenario,
assuming that the full phase-space distribution function is adequately
smoothed we have performed the thermodynamic description of the
non-equilibrium steady states. We have derived a minimum principle for those
stationary states satisfied by the corresponding Lyapounov function which
coincides with the entropy production referred to the steady distribution vector.

Therefore, in our opinion we have reconciled the reversibility of the
microscopic level of description with the irreversible macroscopic behavior
giving a molecular basis to the Non-equilibrium thermodynamics not restricted
to the linear range. This enables us to perform the thermodynamic description
of non-equilibrium steady states that can be far away from equilibrium,
avoiding divergences and possible violations of the second law referred to in
the previous literature. Hence, the answer to the question raised at the
beginning, can Thermodynamics describe systems far from equilibrium?, is
obviously yes if one works at the adequate level of description, which
corresponds to the distribution vector.

\section*{Appendix: local form}

The local form corresponding to Eq. (\ref{entropyproduction3}) is
\begin{gather}
\frac{\partial}{\partial t}f_{1}(x,t)s(x,t)\nonumber\\
=\frac{1}{T}\sum_{n=1}^{N}\frac{1}{n!}\sum_{j=1}^{n}\int f_{n}\underline
{p}_{j}\cdot\left(  -k_{B}T\nabla_{j}\log f_{o,n}+\underline{\mathcal{F}}%
_{j}+\sum_{j\neq i=1}^{n}\underline{F}_{j,i}\right)  \delta(x_{1}%
-x)dx_{1}.....dx_{n}\text{ \ } \label{local}%
\end{gather}
or
\begin{gather}
\frac{\partial}{\partial t}\rho(\underline{q},t)s(\underline{q},t)\nonumber\\
=\frac{1}{T}\sum_{n=1}^{N}\frac{1}{n!}\sum_{j=1}^{n}%
{\displaystyle\int}
f_{n}\underline{p}_{j}\cdot\left(  -k_{B}T\nabla_{j}\log f_{o,n}%
+\underline{\mathcal{F}}_{j}+\sum_{j\neq i=1}^{n}\underline{F}_{j,i}\right)
\delta(x_{1}-x)dx_{1}.....dx_{n}d\underline{p}\text{ ,} \label{local2}%
\end{gather}
where $\rho(\underline{q},t)$ is the density and $s(\underline{q},t)$ the
specific entropy.

\end{document}